\documentclass[conference, a4paper]{IEEEtran}
\IEEEoverridecommandlockouts
\usepackage{cite}
\usepackage{amsmath,amssymb,amsfonts}
\usepackage{algorithmic}
\usepackage{graphicx}
\usepackage{textcomp}
\usepackage{xcolor}
\usepackage{multirow}
\usepackage{multicol}
\usepackage{array}
\usepackage[T1]{fontenc} % for \k command

\usepackage[utf8]{inputenc}
\usepackage[T1]{fontenc}
\usepackage{tabulary}
\usepackage{tabularx}
\usepackage{tabu}
\def\BibTeX{{\rm B\kern-.05em{\sc i\kern-.025em b}\kern-.08em
    T\kern-.1667em\lower.7ex\hbox{E}\kern-.125emX}}
\begin{document}
\bstctlcite{IEEEexample:BSTcontrol}

\title{Emotion Recognition Using Wearables: A Systematic Literature Review -- Work-in-progress\\
% {\footnotesize \textsuperscript{*}Note: Sub-titles are not captured in Xplore and should not be used}
\thanks{This work was partially supported by the National Science Centre, Poland, project no. 2016/21/B/ST6/01463; European Union’s Horizon 2020 research and innovation program under the Marie Skłodowska-Curie grant agreement No. 691152 (RENOIR); the Polish Ministry of Science and Higher Education fund for supporting internationally co-financed projects in 2016–2019 no. 3628/H2020/2016/2; and the statutory funds of the Department of Computational Intelligence, Wroclaw University of Science and Technology.}
}

\author{\IEEEauthorblockN{Stanisław Saganowski\textsuperscript{1,2,*}, Anna Dutkowiak\textsuperscript{1}, Adam Dziadek\textsuperscript{3}, Maciej Dzieżyc\textsuperscript{1,2}, Joanna Komoszyńska\textsuperscript{1}}
\IEEEauthorblockN{Weronika Michalska\textsuperscript{1}, Adam Polak\textsuperscript{4}, Michał Ujma\textsuperscript{3}, Przemysław Kazienko\textsuperscript{1,2}}
\IEEEauthorblockA{
\textit{\textsuperscript{1}Department of Computational Intelligence, Wrocław University of Science and Technology, Wrocław, Poland} \\
\textit{\textsuperscript{2}Faculty of Computer Science and Management, Wrocław University of Science and Technology, Wrocław, Poland} \\
\textit{\textsuperscript{3}Capgemini, Wrocław, Poland} \\
\textit{\textsuperscript{4}Faculty of Electronics, Wrocław University of Science and Technology, Wrocław, Poland} \\
\textsuperscript{*}stanislaw.saganowski@pwr.edu.pl}
}

\maketitle

\begin{abstract}
Wearables like smartwatches or wrist bands equipped with pervasive sensors enable us to monitor our physiological signals. In this study, we address the question whether they can help us to recognize our emotions in our everyday life for ubiquitous computing. Using the systematic literature review, we identified crucial research steps and discussed the main limitations and problems in the domain.
\end{abstract}


%%%%%%%%%%%%%%%%%%%%%%%%%%%%%%%%%
%%%%%%%%%%%%%%%%%%%%%%%%%%%%%%%%%
%
% REFERENCES
%
% add references to the bibliography.bib file
% you can find it in the upper left corner of the project
% use 'BibTeX' format (google scholar offers it when you click 'cite')
% then use \cite{author2019title} code in text to cite the article
%
%%%%%%%%%%%%%%%%%%%%%%%%%%%%%%%%%
%%%%%%%%%%%%%%%%%%%%%%%%%%%%%%%%%
%
%
% use BRITISH-ENGLISH
%
%%%%%%%%%%%%%%%%%%%%%%%%%%%%%%%%%
%%%%%%%%%%%%%%%%%%%%%%%%%%%%%%%%%

\section{Introduction}

Emotions drive most of our decisions \cite{lerner2015emotion}, not only intuitive ones \cite{kahneman2011thinking}, so they directly affect our everyday life. Most of the research on emotion recognition conducted so far focused on participant (subject) reactions evoked by the prepared stimuli in the controlled environment (laboratory setup). Therefore, complex emotion identification in the real-life environment, especially for pervasive computing, remains a significant challenge for this relatively new but promising field of study.

    It is quite difficult to provide a commonly agreed definition of emotion. We should rather consider a set of features that distinguish an emotion from a non-emotion \cite{smith1990emotion}. Overall, \textit{affect} is seen as a neurophysiological state that is consciously accessible but not directed at any specific entity. \textit{Mood} is a lasting and not very intense sensation. Finally, a short, intense, and directed feeling is described as \textit{emotion} \cite{schmidt2019wearable}. Further in this review, we refer to emotions interchangeably with affect. Besides, it is clear that two affective computing topics: stress and emotion recognition have recently become separate research lines \cite{schmidt2018introducing}.
Emotions can be identified and evaluated from different but complementary points of view:

%Emotions can be identified and evaluated from different points of view, which complement each other:
\begin{enumerate}
  \item Subjective perception of the participant (self-assessment of the subject),
  \item Reaction of participant's organism (physiological signals), which is objective but may be contaminated by the individual's body condition and functioning (influenced by drugs or illnesses),
  \item Behavioral signals like facial expressions, voice, specific body movements or even keystroke patterns \cite{kamdar2016prism},
  \item External evaluation made by the subject's peers, e.g. an adult recognizing the state of the child \cite{feng2018wavelet}.
\end{enumerate}

In this paper, we focus on the second perspective - physiological signals that can be gathered using pervasive sensors built into wearable devices like smartwatches, wrist bands, smart rings or headbands. These kinds of devices possess an undoubted advantage. Due to their unobtrusiveness and convenience, they facilitate pervasive computing and context-aware systems by monitoring human affective states in the real-life environment, a.k.a \textit{field studies}. 

Recently, survey studies on emotion recognition have changed their primary focus from the EEG-based solutions \cite{soroush2017review, xu2018learning}, through facial and speech analysis \cite{marechal2019survey, maria2019emotion}, to physiology-oriented \cite{shu2018review, schmidt2019wearable}. In \cite{shu2018review}, all essential aspects of emotion identification (emotion models, stimuli, features, classifiers, etc.) are described, but it does not consider whether the research was conducted in the lab or field setup. Schmidt at al. respect the environment of the surveyed studies and conclude that emotion recognition outside the controlled lab is much more difficult \cite{schmidt2019wearable}. The authors mention about 50 stress and emotion recognition papers (15 of which were field or constrained field studies).
%and provide practical guidelines for designing and applying ecological-momentary-assessment in field studies.

We go a step further and consider only studies that are (or could be) placed in the field, i.e. they use devices and techniques that allow us to conduct the study in everyday situations. Additionally, by following the systematic literature review (SLR) procedure, we cover all relevant research published up to date. We expect SLR to be completed by the date of the conference. It means the extended results will be presented during the workshop. 

\section{Systematic Literature Review}
A great method to summarize existing knowledge in a given domain is \textit{Systematic Literature Review} (SLR). It involves identifying, evaluating and interpreting available research relevant to a certain research question \cite{kitchenham2004procedures}. In our SLR, we are posing the following research question: \textit{Can wearables be used to identify emotions in everyday life supporting pervasive computing?} To refine the number of studies considered in our SLR, we support our question with a set of criteria.

\noindent
Inclusion criteria:
\begin{itemize}
    \item Various emotions identified.
    \item Personal device/wearable used.
    \item At least one physiological signal monitored.
\end{itemize}

\noindent
Exclusion criteria:
\begin{itemize}
        \item Study performed on a population smaller than five subjects.
        \item Single emotion or its levels is considered.
        \item Device is not personal/wearable/portable.
        \item Device or its modules are connected with a cable.
%    \item Device has modules that are interconnected with cables.
\end{itemize}

We used three databases to find articles relevant to our research question: Scopus, Web of Science, and Google Scholar via Publish or Perish. Our search was narrowed down by the following terms: \textit{emotion, affective, wearable, smartwatch, smart device, smart band, personal device, iot, ambient intelligence}. In total, 2,424 papers were found. At the current stage, 1,104 (45\%) articles were carefully reviewed. Only 27 of them (2.5\%) satisfied our research criteria.
%After completing the full SLR, we will retrieve and add new papers published since the initial search.
We have prioritized papers including \textit{emotion recognition} phrase in the title or abstract; therefore, it is likely that the majority of the relevant papers are already identified at this stage.

\section{General Study Design for Emotion Detection}

\begin{figure}[htbp]
\centerline{\includegraphics[width=0.72\columnwidth]{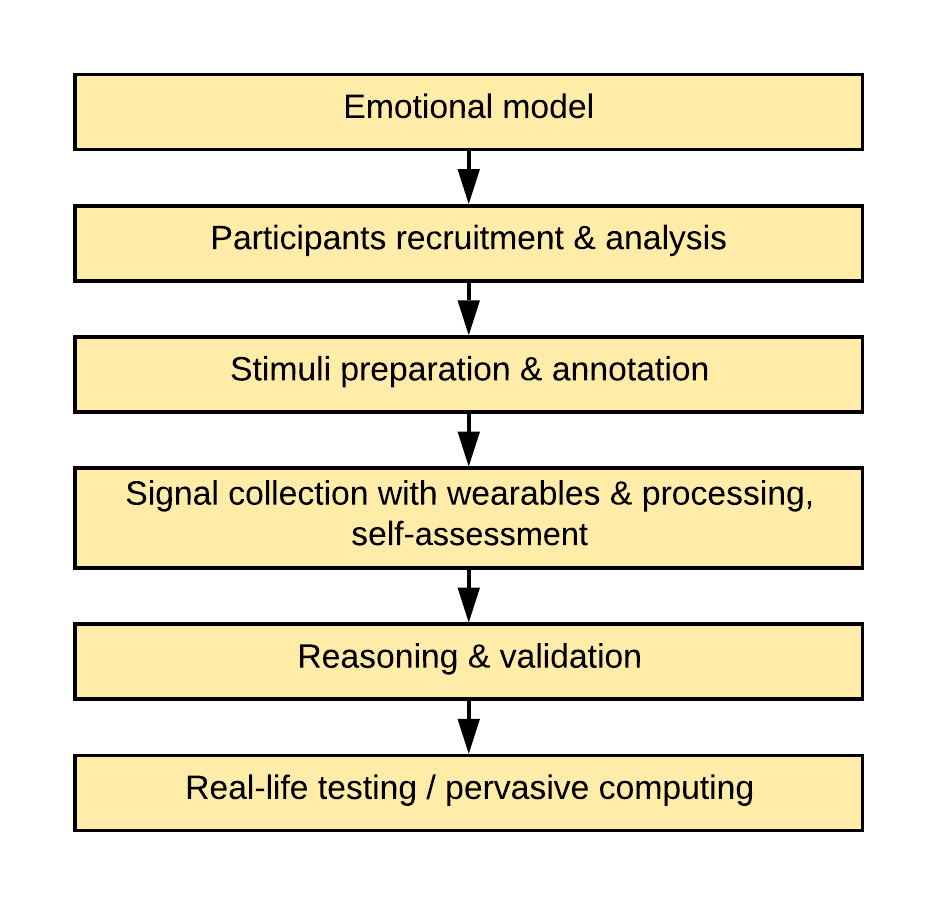}}
\caption{Research stages for emotion recognition in the real-life environment.}
\label{fig:stages}
\end{figure}

\begin{figure}[htbp]
\centerline{\includegraphics[width=0.88\columnwidth]{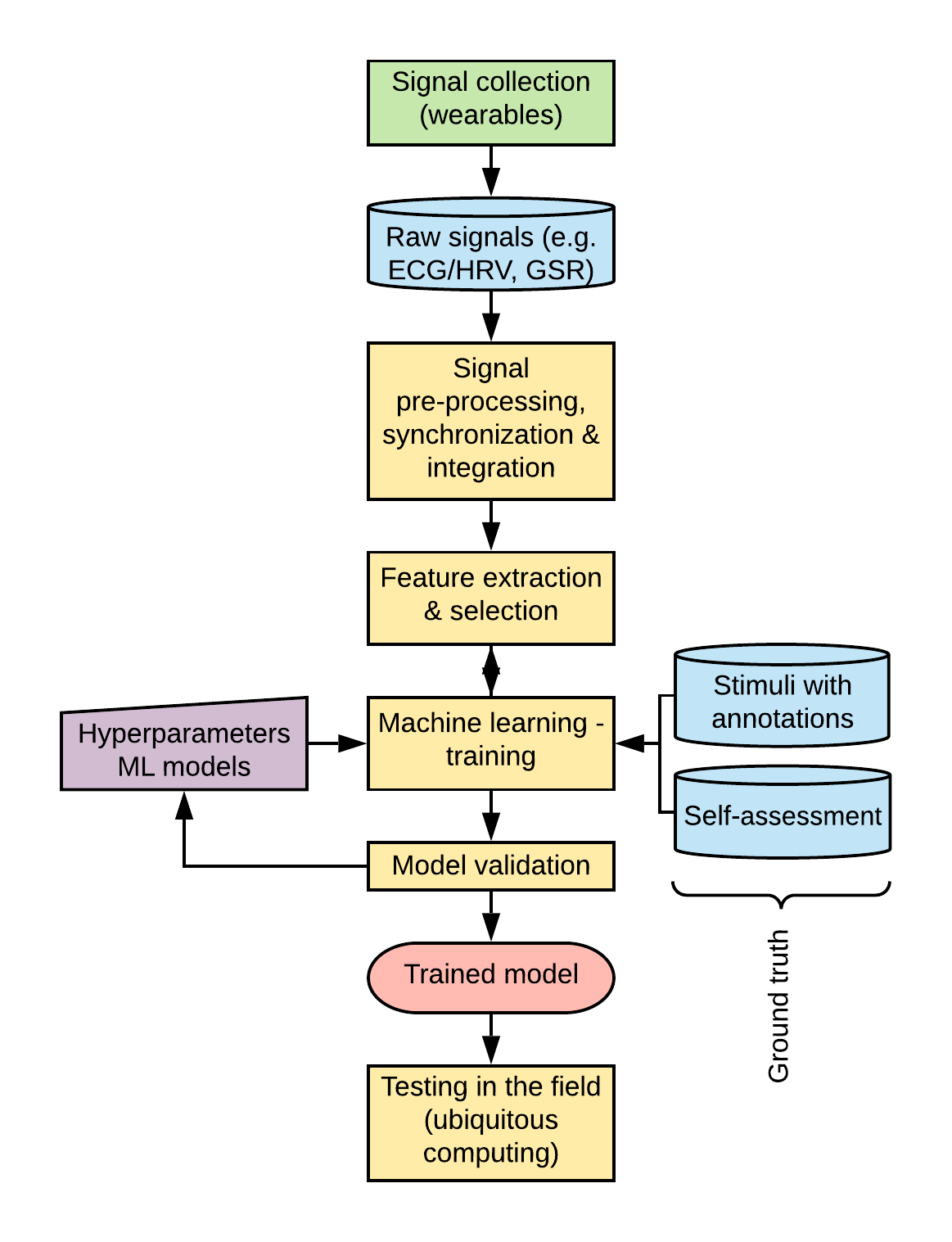}}
\caption{Emotion recognition process using biological signals from wearables.}
\label{fig:emotionDetection}
\end{figure}

We can distinguish six stages in research design for emotion recognition for pervasive computing, Fig. \ref{fig:stages}. Decision about emotional model directly impact on reasoning output, Sec. \ref{sec:emotionalModels}. The second stage covers recruitment, training, profiling and selection of study participants. Some of them need to be excluded due to diseases, e.g. heart problems can significantly interfere with physiological ECG or PPG signals. Then, the appropriate study setup has to be planned differently for a laboratory and field environment (pervasive computing). For laboratory studies, stimuli with annotations may be prepared, whereas ecological momentary assessment (EMA) questionnaires are more suitable for field studies. Next, physiological signals from wearables are collected (Sec. \ref{sec:signals}) together with self-assessments forms. Raw signals are pre-processed, sampled, synchronized and descriptive features are derived, Sec. \ref{sec:preprocessing}, Fig. \ref{fig:emotionDetection}. The selected features combined with the ground truth emotional labels are used to train reasoning model, Sec. \ref{sec:reasoning}. To adjust the model, hyper-parameter optimization should be considered in order to make it ready for ubiquitous computing.
%In order to find the best reasoning model, different machine learning models and hyperparameters should be considered, Fig. \ref{fig:emotionDetection}.

\section{Crucial Research Components}

\subsection{Emotional Models}
\label{sec:emotionalModels}
Since human beings are complex, also their emotions can be modelled using various concepts. There is no single, commonly agreed emotion model. In general, categorical and multidimensional approaches are utilized. The former distinguishes several discrete emotional categories like six types of emotions proposed by Ekman and Friesen \cite{Ekman1978} used in \cite{schmidt2018labelling}; Plutchik's 'wheel of emotions' \cite{Plutchik2003} with 8 primary emotions utilized as 8$\times$3=24 emojis \cite{albraikan2018iaware} (only four+neutral were actually analyzed more in-depth) or limited to only 6 basic emotions \cite{rattanyu2010emotion}. Many authors  decided to apply their own emotion categories, usually as a subset or small modification of the classical ones, e.g. \textit{happy-sad-anger} + neutral state \cite{lu2019sounds} extended in \cite{hu2018scai} with \textit{fear}, \textit{happy-sad-anger-pain} \cite{pollreisz2017simple} \textit{joy-sadness-anger-pleasure} \cite{he2017emotion}, or 
%completely another set:
\textit{anger-sadness-fear-disgust-joy/amusement} + neutral  \cite{rattanyu2011emotion}, also extended with the seventh one -- \textit{affection} \cite{fernandez2019emotion}, \textit{anger-sadness-fear-surprise-frustration-amusement} \cite{lisetti2006categorizing}, extended in \cite{lisetti2004using} with \textit{disgust-other} (they did not applied these two to reasoning), \textit{excited-bored-stressed-relaxed-happy} \cite{dao2018healthyclassroom}. The number of reported categorical emotions can be even greater like as many as eight+neutral in \cite{nakisa2018long}. In \cite{maier2019deepflow}, \textit{happy-sad-flow} categories reflecting flow theory were used.
Sometimes, authors treated each emotion as an independent but potentially co-occurring state, each having its level/scale \cite{lisetti2004using, guo2015short}. It actually makes them a multi-dimensional approach.

A multi-dimensional emotion model assumes multiply dimensions and separate values assigned to each. The most popular is the 2-dimensional \textit{arousal-valence} Russell's model. Arousal denotes affective intensity and valence emotion type (between sad and happy). Typically, dimension values are discrete, e.g. binary states \textit{low-high} \cite{zhao2018emotionsense, nakisa2018long} (forming four quadrants in the 2-dimensional space) or 3-valued \textit{low-medium/neutral-high} each \cite{nalepa2019analysis, ragot2017emotion}. They can be transferred into discrete classes, e.g. \textit{joy-sad-stress-calm} \cite{setiawan2018framework, zhao2018emotionsense}, \textit{happy-bored-neutral} \cite{feng2018wavelet}. Sometimes, the third dimension: \textit{dominance} \cite{xu2019emotion}, \textit{liking} \cite{gupta2016quality} or \textit{relaxed} \cite{kamdar2016prism} is also used. If only valence is applied, we receive a simple binary model like \textit{negative/fear/sad} against \textit{positive/relax/happy} \cite{nie2017emotional}, sometimes extended with neutral \cite{jalilifard2017brain}. Single valence may have multiple levels \cite{kanjo2019deep, jalilifard2017brain}.% The levels, if applied to multiple categorical emotions, actually form a multi-dimensional model \cite{lisetti2004using, guo2015short}.

There are also categorical approaches, which combine stress with some other states like \textit{neutral-stress-amusement} \cite{schmidt2018introducing} or with more other states \cite{dao2018healthyclassroom}.

%\textcolor{red}{PK: do przeniesienia do IV.D self-assesment Commonly, the emotions were identified by means of self-assessment questionnaires filled up by the subjects, except \cite{feng2018wavelet}, in which experts evaluated children's expressions. Alternatively, stimuli were annotated with emotions \cite{lu2019sounds}}.

\subsection{Participants}
\label{sec:participants}

%In SLR, we noticed that there are enormous differences between study conditions and participants preparation before each experiment. Subjects occurring in researches differed in age, gender, nationality and culture. 
%There are significant differences in emotional responding between individuals. It was emphasized that gender differences in any particular modality of emotional expression are culturally and situationally specific \cite{Brody2008}. The stereotype that females are more emotional than men is pervasive across several different cultures \cite{timmers2003ability}. Among North American samples, women are believed to be more emotionally intense \cite{robinson1997emotion}.

There are significant differences in experiencing emotions between individuals. A several studies revealed that the age \cite{fernandez2018emotional}, gender \cite{deng2016gender}, and personality \cite{costa2019emotions} have influence on emotions.
%Only a few studies have considered these factors \cite{}.
In many papers the health of the participants is considered, especially a treatment history and current medications \cite{xu2019emotion,fernandez2019emotion,guo2015short,zhao2018emotionsense,albraikan2018iaware,schmidt2018introducing,kamdar2016prism}, information about vision correction \cite{fernandez2019emotion,ragot2017emotion}, and pregnancy \cite{schmidt2018introducing}. Some experiments are preceded with a questionnaire about depression \cite{fernandez2019emotion,he2017emotion}, excessive sweating \cite{albraikan2018iaware}, or the use of cigarettes/tobacco, alcohol or coffee  \cite{xu2019emotion,ragot2017emotion,zhao2018emotionsense,schmidt2018introducing}.
%Some researches analyzed participants for the probability of depression using the Beck Depression Inventory\cite{fernandez2019emotion,he2017emotion} 
%PHQ-9 Depression Test Questionnaire \cite{zangroniz2017electrodermal}, 
%hyperidrosis \cite{albraikan2018iaware}.
% or psychical or mental disorders with the Maudsley Obsessive Compulsive Questionnaire \cite{fiorini2018physiological} to assess obsessive-compulsive behaviour in subjects.
%Some studies require also information about which hand is dominant [no ref in current review table].
The total number of participants enrolled 
%in the evaluated studies 
vary from two up to nearly two hundred. Some authors disclose whether the study participants volunteered \cite{fernandez2019emotion}, or received a gratification \cite{ragot2017emotion}.

Surprisingly, only a few studies mention they are approved by the ethical committee \cite{fernandez2019emotion,he2017emotion,kanjo2019deep,jalilifard2017brain,albraikan2018iaware}.

\subsection{Signals and Sensors}
\label{sec:signals}

% in the next article apply solution proposed by prof. Polak: Sygnał HR (zazwyczaj to samo co RR), HRV (zazwyczaj to HR) jest zawsze wydobywane z ECG, i czasami analizowany obok „surowego” ECG, jak w art. (19), ale w tabeli tego nie uwzględniono – warto byłoby jeszcze raz przejrzeć artykuły z analizowanym ECG, czy przypadkiem nie są tam analizowane też sygnały pochodne jak HR. Ostatecznie proponuję pozostać tylko przy dwóch sygnałach/skrótach ECG (analiza pełnego sygnału) i HR (zamiast RR, R-R, HRV). Nieelektrycznie z pulsacji tętnic mierzy się BVP i PPG, ale to ten sam sygnał – prościej nazwać go np. blood pulse (BP) – informacja analogiczna do HR. Jedną z reakcji są zmiany w elektrycznych właściwościach skóry. Najbardziej ogólny wydaje mi się termin EDA, zatem warto przemianować na EDA: GSR i EDR (oraz pomyłkowo zapisane GSP w (3) i GRS w (28)). Temperatura rzadko mierzona jest wewnętrznie, więc umieszczone w tabeli pomiary to temperatura ciała / skóry, zatem wspólnie np. body temperature BT, zamiast TEMP, ST, czy SKT. Sygnał oddechowy najczęściej oznaczony jest w tabeli jako RSP, ale czasami też jako RR (repiratory rate) lub IBI (inter-breath interval) – niech zostanie tylko skrót RSP. Akcelerometry (i czasami żyroskop) służą do rejestracji lokalnej aktywności ciała, więc może ogólnie ten sygnał nazwać ACT (niezależnie od metody pomiaru) – przy art. (38) i (45) nie podano, jak była ona mierzona.

Detecting emotions from the physiological state requires monitoring body condition by tracking parameters such as heart rate (HR), galvanic skin response (GSR), body temperature (BT). SLR revealed the following signals (Tab.~\ref{tab:signals}) are used for detecting affect for pervasive computing. Electrocardiography (ECG) measures the electrical activity of the heart, photoplethysmogram (PPG) registers flow changes in blood volume of the monitored vessel, GSR quantifies variation in the skin conductance. Electroencephalography (EEG) records the electrical activity of the brain, and respiration (RSP) is the breathing rate. Additional set of parameters (Tab.~\ref{tab:parameters}) is derived from the above mentioned. HR is the number of heartbeats per minute, while heart rate variability (HRV) describes the variation between interbeat intervals (IBI).

\vspace*{-\baselineskip}
\begin{table}[htbp]
\caption{Physiological signals used for emotion recognition}
\vspace*{-\baselineskip}\begin{center}\vspace*{-\baselineskip}
\begin{tabulary}{\columnwidth}{|L|L|}
\hline
\textbf{Signal} & \textbf{Used by}\\
\hline
ECG & \hspace{1sp}\cite{gupta2016quality, rattanyu2011emotion, guo2015short, he2017emotion, rattanyu2010emotion, hu2018scai}\\
\hline
PPG& \hspace{1sp}\cite{fernandez2019emotion, dao2018healthyclassroom}\\
\hline
GSR, EDA, EDR, SC & \hspace{1sp}\cite{nalepa2019analysis, setiawan2018framework, gupta2016quality, feng2018wavelet, fernandez2019emotion, ragot2017emotion, zhao2018emotionsense, pollreisz2017simple, kanjo2019deep, lisetti2006categorizing, lisetti2004using, albraikan2018iaware, schmidt2018introducing, schmidt2018labelling, dao2018healthyclassroom, maier2019deepflow} \\
\hline
EEG & \hspace{1sp}\cite{lu2019sounds, gupta2016quality, xu2019emotion, nie2017emotional, jalilifard2017brain, nakisa2018long} \\
\hline
RSP & \hspace{1sp}\cite{he2017emotion}  \\
\hline
BT & \hspace{1sp}\cite{zhao2018emotionsense, pollreisz2017simple, kanjo2019deep, lisetti2006categorizing, lisetti2004using, schmidt2018introducing, schmidt2018labelling, dao2018healthyclassroom, maier2019deepflow} \\
\hline
\end{tabulary}
\label{tab:signals}
\end{center}
\end{table}
\vspace*{-\baselineskip} 
% \vspace*{-\baselineskip} % MD: Jak będziemy bardzo zdesperowani to można odkomentować
\vspace*{-\baselineskip}
\begin{table}[htbp]
\caption{Parameters derived from physiological signals}
\vspace*{-\baselineskip}\begin{center}
\begin{tabular}{|l|l|}
\hline
\textbf{Parameter} & \textbf{Used by} \\
\hline
HR & \hspace{1sp}\cite{ragot2017emotion, pollreisz2017simple, kanjo2019deep, lisetti2006categorizing, lisetti2004using, albraikan2018iaware, kamdar2016prism, maier2019deepflow} \\
\hline
HRV & \hspace{1sp}\cite{setiawan2018framework, guo2015short, maier2019deepflow}\\
\hline
IBI & \hspace{1sp}\cite{albraikan2018iaware, kamdar2016prism}\\
\hline
BVP & \hspace{1sp}\cite{nalepa2019analysis, zhao2018emotionsense, schmidt2018introducing, nakisa2018long, schmidt2018labelling} \\
\hline
\end{tabular}
\label{tab:parameters}
\end{center}
\end{table}
\vspace*{-\baselineskip}

We noted several commercially available wearable devices, that seem to be comfortable and useful for collecting physiological signals throughout the day, see Tab.~\ref{tab:devices}.
%Focusing our SLR on the everyday applications means that devices used for physiological signal collection should be wearable and wireless. We noticed a small group of devices used by researchers. All commercially available solutions we encountered are listed in Tab.~\ref{tab:devices}.
In addition, some articles propose self-made devices \cite{rattanyu2011emotion, he2017emotion, rattanyu2010emotion, hu2018scai}.  

\vspace*{-\baselineskip}
\begin{table}[htbp]
\caption{Wearables for collecting physiological signals}
\vspace*{-\baselineskip}\begin{center}
\begin{tabu}{|X[2,l]|X[1,l]|X[2,l]|}
\hline
\textbf{Device} & \textbf{Sensors} & \textbf{Used by}\\
\hline
Empatica E4 & PPG, GSR, BT & \hspace{1sp}\cite{nalepa2019analysis, ragot2017emotion, zhao2018emotionsense, pollreisz2017simple, albraikan2018iaware, schmidt2018introducing, nakisa2018long, schmidt2018labelling, dao2018healthyclassroom, maier2019deepflow} \\
\hline
Microsoft Band 2 & PPG, GSR, BT & \hspace{1sp}\cite{nalepa2019analysis, setiawan2018framework, kanjo2019deep} \\
\hline
Samsung Gear S & HR & \hspace{1sp}\cite{kamdar2016prism} \\
\hline
BodyMedia SenseWear Armband & HR, GSR, BT & \hspace{1sp}\cite{lisetti2004using,lisetti2006categorizing} \\
\hline
Neurosky MindWave & EEG & \hspace{1sp}\cite{lu2019sounds,nie2017emotional, jalilifard2017brain, gupta2016quality} \\
\hline
XYZlife Bio-Clothing & ECG & \hspace{1sp}\cite{guo2015short} \\
\hline
\end{tabu}
\label{tab:devices}
\end{center}
\end{table}
\vspace*{-\baselineskip}
\vspace*{-\baselineskip}

\subsection{Emotion Self-assessment}
% MD: Sprawdzic czy sama kolumna jest poprawna. Na razie nie sprawdzalem tego
% MD: Sprawdzone

A big challenge for studies on human emotions is to obtain ground truth. Authors most commonly request subjects to self-report their emotional state. However, no information or not enough details about self-assessment are the issue in seven studies \cite{nalepa2019analysis, setiawan2018framework, feng2018wavelet, fernandez2019emotion, nie2017emotional, he2017emotion, hu2018scai}.

%Others applied some kind of self-assessment. 
Authors usually ask about the intensity of the experienced emotions in Likert scale (five papers) \cite{zhao2018emotionsense, guo2015short, gupta2016quality, jalilifard2017brain, kamdar2016prism} or the subjects select their emotions from the closed list (three cases) \cite{schmidt2018labelling, nakisa2018long, dao2018healthyclassroom}. In \cite{lu2019sounds}, the open-ended questions are used, and in two other papers, the combination of open-ended and close-ended questions is applied \cite{lisetti2006categorizing, lisetti2004using}.

Some researchers use the well-established questionnaires or their modified versions. In six studies, Self Assessment Manikin (SAM) is applied \cite{xu2019emotion, ragot2017emotion, kanjo2019deep, pollreisz2017simple, schmidt2018labelling, schmidt2018introducing}. Much less frequently, the IAPS qustionare is utilized \cite{rattanyu2011emotion, rattanyu2010emotion}, AniAvatare \cite{albraikan2018iaware}, Photographic Affect Meter (PAM) \cite{schmidt2018labelling}, State-Trait Anxiety Inventory (STAI) \cite{schmidt2018labelling}, Game Experience Questionnaire (GEQ) \cite{maier2019deepflow}, Positive and Negative Affect Schedule (PANAS) \cite{schmidt2018introducing}, and Short Stress State Questionnaire \cite{schmidt2018introducing}. In  \cite{feng2018wavelet}, adult experts assess emotional state based on children face expression.

\subsection{Stimuli}
\label{sec:stimuli}
There are many procedures to elicit emotions. Most often used are affective videos \cite{fernandez2019emotion,guo2015short,he2017emotion,zhao2018emotionsense,pollreisz2017simple,lisetti2006categorizing,jalilifard2017brain,albraikan2018iaware,schmidt2018introducing,nakisa2018long}, images \cite{nalepa2019analysis,nie2017emotional,ragot2017emotion,rattanyu2011emotion,rattanyu2010emotion}, and music/sounds \cite{lu2019sounds,nalepa2019analysis,gupta2016quality,he2017emotion,magno2017deepemote}. They are easy to explain, simple to apply and available in many data sets: IAPS \cite{lang2005iaps}, NAPS \cite{marchewka2014naps}, CAPS \cite{lu2005caps}, IADS \cite{bradley2007iads}, MAHNOB \cite{soleymani2011mahnob}.

On the other hand, many studies use stimuli that can be experienced in real-life like playing a game \cite{nalepa2019analysis,maier2019deepflow}, solving a math problem \cite{lisetti2006categorizing}, learning \cite{dao2018healthyclassroom}, walking around the city \cite{kanjo2019deep}, and other \cite{schmidt2018labelling,kamdar2016prism,feng2018wavelet,sokolova2015distributed}. However, most of the experiments were performed in the lab, where obtaining high-quality measurements is relatively easy, and only a few studies were conducted in an uncontrolled environment \cite{kanjo2019deep,kamdar2016prism,schmidt2018labelling,albraikan2018iaware}.

Usually, the exposure to stimuli lasts 5-10 seconds per image, 15-30 seconds per audio, 2-5 minutes per video, and 2-60 minutes in case of everyday activities.

%\subsection{MD: Data Sets}

%Currently we have identified only one publicly available dataset which was a result of a study which meets all inclusion criteria and does not meet any exclusion criteria (especially used device is wearable, and at least two emotions are identified), i.e. Multimodal Dataset for Wearable Stress and Affect Detection (WESAD). WESAD consists of data of 15 subjects which were subjected to amusement and stress condition and were wearing Empatica E4 and RespiBAN Professional \cite{schmidt2018introducing}.

\subsection{Signal Processing}
\label{sec:preprocessing}

Most of the papers follow the classical approach to signal processing in the field of machine learning, consisting of three explicit stages: pre-processing, feature extraction and selection, Fig. \ref{fig:emotionDetection}. The role of pre-processing is to remove from signals information that is not related to emotional patterns and can negatively impact on results. It also helps to extract the discriminative features more efficiently. The most frequently used pre-processing methods are gathered in Table~\ref{tab:signal_preprocessing}.

\vspace*{-\baselineskip}
\begin{table}[htbp]
\caption{The most popular methods used for pre-processing}
\vspace*{-\baselineskip}\begin{center}
\begin{tabu}{|X[3,l]|X[2,l]|}
\hline
\textbf{Pre-processing method} & \textbf{Used by} \\
\hline
Filtration (lowpass, bandpass, notch, median, drift removal) & \hspace{1sp}\cite{feng2018wavelet, fernandez2019emotion, xu2019emotion, guo2015short, zhao2018emotionsense, schmidt2018introducing, nakisa2018long, hu2018scai, schmidt2018labelling, dao2018healthyclassroom} \\
\hline
Normalization (to ±1 or [0,1] range) and standardization (dividing by SD) & \hspace{1sp}\cite{setiawan2018framework, nie2017emotional, lisetti2006categorizing, lisetti2004using, dao2018healthyclassroom, maier2019deepflow}\\
\hline
Winsonization (removing outliers and dubious or corrupted fragments, interpolation of removed samples) & \hspace{1sp}\cite{feng2018wavelet, he2017emotion}\\
\hline
\end{tabu}
\label{tab:signal_preprocessing}
\end{center}
\end{table}
\vspace*{-\baselineskip}

The stage of feature extraction is used to reduce the dimensionality of the problem while maintaining the relevant information. It results in the feature vector representing the original signal or its segment. Usually, features are calculated in time or other domain Tab.~\ref{tab:signal_feature_extraction}, i.e. an original variable is transformed or decomposed into other signals. Then, different metrics are superimposed on them. 
%The most commonly applied groups of methods are summarized in Tab.~\ref{tab:signal_feature_extraction}.

The last stage, consisting in the selection of a subset of features, focuses on the further reduction of dimensionality, taking into account the redundancy of previously extracted features or their inability to distinguish between considered classes. Carrying it out increases the efficiency of classification algorithms. Despite this, the feature selection is not mentioned or performed in most of the reviewed papers. Only in \cite{he2017emotion} a genetic algorithm is used, and in \cite{zhao2018emotionsense} Sequential Forward Floating Selection is applied. Two more works perform Principal Component Analysis (PCA) \cite{feng2018wavelet, zhao2018emotionsense}.

\subsection{Reasoning Models}
\label{sec:reasoning}

% TODO: for camera-ready version rewrite this section focusing on binary / multi-class / multi-label classification
%Multiclass problem is often solved by changing to binary form. Possible approaches are ... \textcolor{red}{xxxx???}

Only four studies apply a deep learning algorithm such as convolutional neural networks \cite{maier2019deepflow,kanjo2019deep}, long short-term memory neural networks \cite{nakisa2018long,kanjo2019deep}, or neural networks with backpropagation \cite{lisetti2006categorizing}. All other works use simple supervised learning models among which Support Vector Machine is the most often used.

\vspace*{-\baselineskip}
\begin{table}[htbp]
\caption{The most popular methods used for feature extraction}
\vspace*{-\baselineskip}\begin{center}
\begin{tabu}{|X[18,l]|X[33,l]|X[24,l]|}
\hline
\multicolumn{2}{|c|}{\textbf{Feature extraction method}} & \textbf{Used by} \\
\hline
\multirow{2}{*}{Time domain} & Signal morphology (amplitude, extrema, intervals, etc.) & \hspace{1sp}\cite{nalepa2019analysis, fernandez2019emotion, xu2019emotion, rattanyu2011emotion, he2017emotion, pollreisz2017simple, lisetti2006categorizing, lisetti2004using, schmidt2018introducing, maier2019deepflow, kamdar2016prism, hu2018scai} \\ \cline{2-3}
 & Rate of specific events & \hspace{1sp}\cite{ragot2017emotion, rattanyu2011emotion, guo2015short, he2017emotion, schmidt2018introducing, kamdar2016prism} \\ \cline{2-3}
 & RMS & \hspace{1sp}\cite{nalepa2019analysis, fernandez2019emotion, rattanyu2010emotion} \\ \cline{2-3}
\hline
\multirow{2}{2 cm}{Frequency domain} & PSD & \hspace{1sp}\cite{gupta2016quality, xu2019emotion, ragot2017emotion, zhao2018emotionsense, schmidt2018introducing, nakisa2018long} \\ \cline{2-3}
 & Frequency spectrum & \hspace{1sp}\cite{gupta2016quality, guo2015short, zhao2018emotionsense, nakisa2018long} \\ \cline{2-3}
\hline
Time-frequency or time-scale domain & STFT, WT (CWT, DWT, FWT) & \hspace{1sp}\cite{feng2018wavelet, xu2019emotion, rattanyu2010emotion, jalilifard2017brain} \\
\hline
Statistical indices & Mean, median, SD, skewness, kurtosis, etc. & \hspace{1sp}\cite{nalepa2019analysis, setiawan2018framework, gupta2016quality, feng2018wavelet, fernandez2019emotion, xu2019emotion, ragot2017emotion, rattanyu2011emotion, guo2015short, he2017emotion, zhao2018emotionsense, rattanyu2010emotion, lisetti2006categorizing, lisetti2004using, schmidt2018introducing, nakisa2018long, kamdar2016prism} \\
\hline
Nonlinear measures & Measures of chaos, complexity, entropy & \hspace{1sp}\cite{xu2019emotion, zhao2018emotionsense, nakisa2018long} \\
\hline
\end{tabu}
\label{tab:signal_feature_extraction}
\end{center}
\end{table}
\vspace*{-\baselineskip}

%some studies performed the machine learning part unreliably, e.g., Lisseti et al. \cite{lisetti2006categorizing} used Marquardt Backpropagation Algorithm \cite{hagan1994mba} which uses the sum of squares error in the performance function, meanwhile authors report the results using accuracy. Also, using the accuracy measure when data is imbalanced.

Really surprising is the fact that only two papers mention imbalance in collected samples \cite{gupta2016quality,maier2019deepflow}.
% survey by Bosch also doesn't mention this problem!!!
When dealing with multiple emotions (or even just two) it is very unlikely that the distribution will be equal, i.e. each emotion will occur a similar number of times. At the same time, the majority of works use the accuracy as the classification quality measure, which is not appropriate for imbalanced data sets (unless calculated for the correctly classified cases only).

Furthermore, most papers do not provide information on model validation, in particular, evaluated hyper-parameters. Just three compare diverse setups \cite{feng2018wavelet, he2017emotion, nakisa2018long}, two examine different feature sets \cite{rattanyu2011emotion, zhao2018emotionsense}, and a few test various algorithms \cite{nalepa2019analysis, xu2019emotion, rattanyu2011emotion, zhao2018emotionsense, lisetti2006categorizing, lisetti2004using, kamdar2016prism}. Only \cite{kanjo2019deep} addresses a vital question whether to train a general model for all subjects or to build multiple personalized classifiers adjusted to each individual.

The most common validation techniques are k-fold and leave-one-out cross-validation. Labels (output and ground truth for models) are usually obtained from self-assessments or annotated stimuli data. A less popular approach is to employ experts (psychologists) \cite{feng2018wavelet,gupta2016quality,jalilifard2017brain}.% or the external system \cite{albraikan2018iaware}.
%Usage of multiple label sources causes conflicts when participant perception differs from stimulus annotations. Such cases can be resolved by defining only one ground truth source. Alternatively, ambiguous data can be removed from the data set \textcolor{red}{PK: potrzebne odeslanie do literatury}.

%Interesting usage of machine learning is checking the signal quality before further analysis \cite{gupta2016quality}.

%Something about evaluations metrics
%something about data split

\section{Applications to Pervasive Computing and Context-aware Systems}

Affect recognition presents a wide range of possible applications in pervasive computing. With progressive advancement in physiological sensor signal quality and miniaturization, use cases including emotion detection are at the fingertips. During SLR, we have noticed two main application areas emerging: human-computer interaction (HCI) and healthcare.

HCI can be greatly improved once the computer understands human affect \cite{ setiawan2018framework}. An affective loop can be established to learn and foster user experience, and virtual assistants may respond better \cite{nalepa2019analysis}. Recommendations for
%movie \cite{kutt2018bandreader}, music \cite{kutt2018bandreader},
search engine results \cite{gupta2016quality}, user interface and content \cite{lisetti2004using, lisetti2006categorizing} may be enhanced with user's emotional context. Affect-aware robots might provide better user experience \cite{rattanyu2010emotion, rattanyu2011emotion}, and television content suggestions could be more accurate \cite{jalilifard2017brain}. Emotions play an important role in computer games, \cite{nalepa2019analysis} supporting game-world design process based on a player's affective model. Player's experience can be crafted to emotional feedback \cite{xu2019emotion}, making it more realistic.  

Healthcare can benefit from the fact that the physiological system and ability to feel are intertwined. Ubiquitous emotion detection can aid to monitor our well-being  \cite{hu2018scai, fernandez2019emotion}.
%One of the means for monitoring our well-being could be emotion detection system \cite{hu2018scai, fernandez2019emotion}.
%; mental health is an important factor of well-being.
Furthermore, emotion recognition could help at stress control, e.g. in order to reduce the probability of cardiovascular diseases 
\cite{schmidt2018introducing}. 
%\cite{schmidt2018introducing, gjoreski2017monitoring}. 
Mental health could be monitored by means of affect detection 
\cite{guo2015short, kamdar2016prism, kanjo2019deep}; emotional self-awareness can improve mental health \cite{albraikan2018iaware}
%\cite{guo2015short, zangroniz2017electrodermal,tacconi2008activity, kamdar2016prism, kanjo2019deep, gjoreski2017monitoring}.
In \cite{zhao2018emotionsense}, the authors alert that negative emotional states can have a degrading impact on our health.
Therapeutic contexts are proposed in \cite{pollreisz2017simple,feng2018wavelet}, where children with autism could be taught to understand their emotions better. Affect recognition is also proposed as a comprehensive tool for helping people with emotion-based disorders \cite{he2017emotion}.

Other applications include: detecting driver drowsiness and establishing cognitive load\cite{nakisa2018long}, monitoring classroom attitude\cite{dao2018healthyclassroom}, lesson content difficulty adjustment \cite{maier2019deepflow}, and creating alternative emotional channel via music \cite{lu2019sounds}.

\section{Discussion}

% https://userpages.uni-koblenz.de/~laemmel/esecourse/slides/slr.pdf

% Quality assessment needs to be done - scores?

% Discussion about employed procedure 

%textcolor{red}{discussion on exclusion criteria - should we include smartphones or not? (criteria about physiological signal)}

Simple emotional models like \textit{low arousal-high arousal} or \textit{no stress-low stress-high stress} were extensively explored. It mainly resulted from the strong correlation between arousal or stress and some biological signals - GSR or BVP \cite{schmidt2019wearable, choi2012development}. However, emotions are much more complex \cite{Cowen2017}, and their multidimensional nature remains a great challenge for future work. This also appears to be the main reason why in only a few studies, the authors monitor and recognize emotions in the non-controlled (field study) \cite{kanjo2019deep,kamdar2016prism,schmidt2018labelling} or semi-controlled environment, e.g. emotion monitoring while watching the live football game in TV \cite{albraikan2018iaware}.

%Only few studies were able to monitor and recognize emotions in the non-controlled (field study) \cite{schmidt2018labelling} or semi-controlled environment, e.g. emotion monitoring while real-life watching football matches in TV \cite{albraikan2018iaware}.

Emotion model directly impacts on the detection model. Various emotion models make it almost impossible to compare results with each other, e.g. does \textit{fear} from \cite{hu2018scai} corresponds to \textit{pain} in \cite{pollreisz2017simple}? Multi-dimensional models should straightforward lead to a multi-label classification problem, which in turn requires much more cases to train the classifier. None of the paper approach to reasoning in this way. Moreover, correct multi-label models can recognize combinations that do not occur in the training set.

%None considered relationships between categorical or dimensional emotions, which may result in emotional patterns \cite{Cowen2017}}

There is actually no research, which could be seen as comprehensive with respect to all research design stages, Fig.~\ref{fig:stages}. For example, only two out of all 27 papers consider and properly solve the imbalanced data problem, three optimize machine learning model parameters, five claim to have an ethical committee approval, another five recruited more than 50 participants, sampling rates of the sensor signals are commonly not considered at all. Finally, only four identify complex emotions in real-life conditions.

%\textcolor{red}{indywidualizacja}

We focused on pervasive wireless wearables that can monitor physiological signals. However, some other signals and data may be complementary, e.g. data gathered by our smartphone about our activity \cite{sasaki2018comparing}, voice \cite{denman2018affsens}, or smartwatch built-in camera monitoring our face \cite{rincon2018intelligent}.

The low variety of hardware tested might stem from the fact that most off-the-shelf wearables do not provide access to raw physiological signals.

%do nastepnego artykulu: A lot of research incorrectly mixes up PPG signal and labels with ECG. They are completely different signals based on separate physiological phenomena.   

%Niektore badania powielaja bledy, e.g. no disgust translation to Polish \textcolor{red}{To bym uzyl w drugim artykule - jest malo wazne(do komentarza)}

\section{Conclusions}

In our systematic literature review (SLR), we analyzed methods and solutions that enable us to effectively recognize human emotions in everyday life and utilize such information in pervasive computing. Half-way through, based on the literature, we believe that such a solution is achievable but still requires further investigation.

There are still many challenges and problems to solve in order to ensure high quality of emotion detection in context-aware applications. We found a great potential, especially in the improvement of signal data processing, model learning, and tailoring the suitable deep machine learning architecture. 

We have shown that emotion recognition from personal devices is already a mature but very promising direction for further research.

%czesc odpowiedzialna za machine learning wydaje sie jak na razie najslabiej eksplorowanym polem. tylko dwie prace uzyly LSTM/GRU or some other architectures (e.g. with attention), ktory wydaje sie najbardziej sensowny.  \textcolor{red}{chyba jednak bym tego na razie 'nie sprzedawal'}

%\section*{Acknowledgment}

%To remove The preferred spelling of the word ``acknowledgment'' in America is without an ``e'' after the ``g''. Avoid the stilted expression ``one of us (R. B. G.) thanks $\ldots$''. Instead, try ``R. B. G. thanks$\ldots$''. Put sponsor acknowledgments in the unnumbered footnote on the first page.

\bibliography{bibliography}
\bibliographystyle{IEEEtran}

\end{document}